\documentclass[amsmath,aps,a4paper,nofootinbib]{revtex4}

 \newcommand{\lap}{\bigtriangleup\,}
 \newcommand{\eps}{\varepsilon}
 \newcommand{\bbbr}{\mathrm{I\!R}}
 \newcommand{\tr}{\mathrm{Tr}}
 \newcommand{\ro}{\varrho}

\begin{document}

  \title{MICROSCOPIC THEORY OF THE CASIMIR EFFECT} 
  \date{\today}
  \author{Luca Valeri}
  \author{G\"unter Scharf}
  \email[e-mail:]{scharf@physik.unizh.ch\\}
  \affiliation{ Institut f\"ur Theoretische Physik, \\ 
             Universit\"at Z\"urich, \\ 
             Winterthurerstr.190 , CH-8057 Z\"urich, Switzerland}

 
\begin{abstract} 
Based on the photon-exciton Hamiltonian a microscopic
theory of the Casimir problem for dielectrics is developed. Using
well-known many-body techniques we derive a perturbation expansion for the
energy which is free from divergences. In the continuum limit we turn off the
interaction at a distance smaller than a cut-off distance $a$ to keep the
energy finite. 
We will show that the macroscopic theory of the Casimir
effect with hard boundary conditions is not well defined because it ignores
the finite distance between the atoms, hence is including infinite self-energy
contributions. Nevertheless for disconnected bodies the latter do not 
contribute to
the force between the bodies. The Lorentz-Lorenz relation for the dielectric
constant that enters the force is deduced in our microscopic theory without
further assumptions. 

The photon Green's function can be
calculated from a Dyson type integral equation. The geometry of
the problem only enters in this equation through the region of integration 
which is equal to the region occupied by the dielectric. The
integral equation can be solved exactly for various plain and spherical
geometries without using boundary conditions.
This clearly shows that the Casimir force for dielectrics is due to the
forces between the atoms.

Convergence of the perturbation expansion and the metallic limit are
discussed. We conclude that for any dielectric function the transverse
electric (TE) mode does not contribute to the zero-frequency term of the
Casimir force. 
\end{abstract} 
  
\maketitle

\section{Introduction}


More than 50 years ago Casimir published two completely different papers
\cite{casimir1} \cite{casimir2} on what was later called the Casimir effect: 
One \cite{casimir1} was a cumbersome microscopic theory, the other 
\cite{casimir2} a very elegant macroscopic one. 
From the latter one may get the impression that the problem is
rather simple, namely, quantization of the free electromagnetic field in
a region with boundaries. However, the results of such macroscopic
calculations (the so-called mode summation method) are notoriously
divergent (\cite{bordag} and the many references therein). Although by 
suitable regularizations one may get
reasonable results, the ultraviolet divergences cannot be eliminated in
general. This shows that the macroscopic problem (with hard boundary
conditions) is not well-posed. A second weak point of the macroscopic
method is that it completely ignores the fact that the real Casimir
force is the result of electromagnetic interaction between neutral
atoms and not a free problem. This was already the point of view of
Casimir and Polder in their first paper \cite{casimir1}.
It is clear from the above considerations that a real understanding of
the Casimir problem can only come from a microscopic theory. 

In the following we develop such a theory for dielectrics and use it to solve
various Casimir problems explicitly.  
The basis is the photon-exciton Hamiltonian given
in the next section. The interaction consists of the non-relativistic 
dipole coupling to the
radiation field and the dipole-dipole approximation of the Coulomb
interaction.
In Section \ref{sec:energy} we derive the perturbation
expansion for the energy $E$ for arbitrarily spaced atoms. This energy is finite
since infinite self-energy contributions of the dipoles are excluded. 

When taking the limit to a
continuous dipole distribution one has to be carful because of the
dipole-dipole approximation. We turn off the interaction at an average
distance $a$ between the dipoles. This was also done by Marachevsky
\cite{marachevsky}, leading to a finite energy for a dilute ball.
To extract the classical Casimir energy from $E$ we have to introduce a
further ultraviolet cut-off $e^{-\delta k}$. $E^\delta$ splits then into two
parts: $E^\delta=E^\delta_D-E^\delta_G$. In the limit $\delta\to 0$
only$E^\delta$ remains finite, whereas $E^\delta_D$ and $E^\delta_G$ are
divergent. In the limit $a\to 0$ also $E$ is divergent. $E_D^\delta$ is for
$\delta,a\to0$ formally equivalent to the usual Casimir energy $E_D$ as given
by Candelas \cite{candelas}. It is always infinite.

To calculate the force between separated bodies in regions $A$ and $B$,
one has to use the free energy $F$. It splits into three finite parts: 
$F=F_A+F_B+F_{AB}$. $F_A$ and $F_B$ contain the free energies of the two isolated
bodies. They are finite since the finite distance between the atoms is taken
into account and self-energies are excluded. 
Only the interaction energy $F_{AB}$ depends on the distance of the
bodies and contributes to the force. In the limit $a\to0$ only this energy remains
finite. For the polarization we obtain the Lorentz-Lorenz relation without
further assumptions.
The force between disconnected bodies can be calculated by differentiating $F$
with respect to the distance or by a contour integral 
over a region containing one body. Only the stress field outside the
bodies contributes to the force and is finite also when the cut-off distance
goes to $0$.  

In Section \ref{sec:basic} the Green's function of the total electromagnetic
field for a continuous dipole distribution field is discussed. It is given by
a Dyson type integral equation. 
In this equation the many-body problem is separated from the
geometry. The latter only enters through the region of integration which
agrees with the region of space occupied by the dielectric. We
identify a class of geometries which allow exact solution of the
integral equation. As applications we consider in Section \ref{sec:plain} and
\ref{sec:force} plain  
geometries and we re-derive the Lifshitz formula for the Casimir force
between two plates \cite{lifshits}.
In Section \ref{sec:spherical} we solve the integral equation for the 
Green's function of the
electromagnetic field for a dielectric ball.

In Section \ref{sec:metallic} the metallic limit $\eps\to\infty$ is
considered. Within our theory because of the Lorentz-Lorenz relation 
this means to put the atomic polarization
$\alpha_0(u)=3$ for $u=0$. It is not clear whether that the perturbation 
expansion
converges. We'll show convergence in flat geometries just for $u=0$ 
in the TM mode and for small
frequencies $u>0$ in the TE mode. 
The TE mode does not contribute in the zero-frequency case $u=0$ to the 
Casimir force independent of the model adopted for the dielectric function.
This result cannot be deduced from the Lifshitz formula 
for the plasma model where
$\lim_{u\to0}\eps(iu)u^2\neq0$. The zero-frequency behavior
plays an important role in the context of the temperature dependence of the
Casimir force and is widely discussed in recent papers
\cite{temp0,temp1,temp2,temp3,temp4,temp5,temp6}. 


\section{\label{sec:model}The model}


We consider $N$ atoms in interaction with the quantized radiation field.
The dynamical degrees of freedom of the atoms are their internal
excitations, therefore, our system is described by the following photon
- exciton Hamiltonian:
\begin{equation}
  \label{eq:2.1}
  H=H_0^a+H_0^{ph}+H_1,
\end{equation}
\begin{equation}
  \label{eq:2.2}
  H_0^a=\sum_{\vec x,n}E_n b_n^+(\vec x)b_n(\vec x),
\end{equation}
\begin{eqnarray}
  \label{eq:2.3}
  H_0^{ph}&=&\frac{1}{2}\int d^3x
    \Bigl(:\vec E(\vec x)^2:+:\vec B(\vec x)^2:\Bigl)\nonumber\\
  &=&\sum_{\lambda=1,2}\int d^3x\,ka_\lambda^+(\vec k)a_\lambda(\vec k)
\end{eqnarray}
\begin{eqnarray}
  \label{eq:2.4}
  H_1&=&\sum_{\vec x}\Bigl[-\frac{e}{mc}\vec p_{\vec x}\vec A(\vec x)+
  \frac{e^2}{ 2mc^2}:\vec A(\vec x)^2:\Bigl]\nonumber\\
  &&-\frac{1}{2}\sum_{\vec x\ne
    \vec x\,'}\Phi(\vec x-\vec x\,').
\end{eqnarray}
The atoms sit at places $\vec x_j$,
but for simplicity of the notation we omit the index $j$ and write the
sum over all atoms as $\sum_{\vec x}$. $b_n^+(\vec x)$ and $b_n(\vec x)$
are emission and absorption operators for an atom in state $n$ at the
place $\vec x$. The interaction $H_1$ consists of
the usual dipole coupling of non-relativistic radiation theory plus the
dipole - dipole interaction $\Phi$. We assume that the distance between
the atoms is big compared with the extension of the atoms (Bohr radius).
Then $\Phi$ is given by
\begin{equation}
  \label{eq:2.5}
  \Phi(\vec x-\vec x\,')=\sum_{i,j=1}^3 q_i(\vec x)q_j(\vec x\,')\Phi_{i,j}
  (\vec x-\vec x\,')
\end{equation}
\begin{equation}
  \label{eq:2.6}
  \Phi_{i,j}(\vec R)=\frac{1}{4\pi R^3}
    \left(\delta_{ij}-3\frac{R_iR_j}{R^2}\right),
\end{equation}
where $\vec q(\vec x)$ denotes the dipole operator of an atom at place $\vec x$.
It can be expressed by one-particle matrix elements of the position
operator, similarly to the momentum operator $\vec p(\vec x)$:
\begin{equation}
  \label{eq:2.7}
  \vec p(\vec x)=\sum_n (\varphi_0,\vec p\varphi_n)b_n(\vec x)+
  (\varphi_n,\vec p\varphi_0)b^+_n(\vec x).
\end{equation}
The commutation relations for the emission and absorption operators are
\begin{equation}
  \label{eq:2.8}
  [b_m(\vec x), b_n^+(\vec x\,')]=\delta_{mn}\delta_{\vec x\vec x\,'},
\end{equation}
and zero otherwise. $\delta_{\vec x\vec x\,'}$ is actually a Kronecker
delta, it is =1 if $x'=x$ and 0 otherwise. The quantized electromagnetic
field is given in the radiation gauge by
\begin{equation}
  \label{eq:2.9}
  \vec A(\vec x)=\frac{1}{(2\pi)^{3/2}}\sum_{\lambda=1,2}\int \frac{d^3k}{\sqrt{2k}}
  \vec\xi_\lambda(\vec k)
  \left(a_\lambda(\vec k)e^{i\vec k\vec x}
       +a_\lambda^+(\vec k)e^{-i\vec k\vec x} \right),
\end{equation}
where $\vec\xi_\lambda(\vec k)$ are transversal polarization vectors:
\begin{equation}
  (\vec\xi_\lambda(\vec k),\vec\xi_{\lambda'}(\vec k))=\delta_{\lambda
    \lambda'},
\end{equation}
\begin{equation}
  \label{eq:2.10}
  (\vec\xi_\lambda(\vec k),\vec k)=0.
\end{equation}
The photon emission and absorption operators satisfy
\begin{equation}
  \label{eq:2.11}
  [a_\lambda(\vec k),a^+_{\lambda'}(\vec k\,')]=\delta_{\lambda\lambda'}
  \delta(\vec k-\vec k\,')
\end{equation}
and zero otherwise. 


\section{\label{sec:energy}Energy and the stress tensor}


We use the Matsubara formalism to calculate the expectation value for the
energy for finite temperature.
The expectation value of Hamiltonian is given by:
\begin{equation}
  \label{eq:3.1}
  E=\langle H \rangle=\frac{\tr(e^{-\beta H}H)}{\tr(e^{-\beta H})}
   =-\partial_\beta \ln \tr (e^{-\beta H}) 
\end{equation}
With 
\begin{equation} 
  S=e^{-\beta H_0}\,e^{-\beta H} \label{eq:3.2}
\end{equation}
we obtain
\begin{eqnarray}
  E&=& {}-\partial_\beta \ln\frac{\tr ( e^{-\beta H_0}\,S )}{\tr(e^{-\beta H_0})}
  - \partial_\beta \ln \tr(e^{-\beta H_0})\nonumber\\
  &=& {}-\partial_\beta\ln\langle S\rangle_0+ \langle H_0\rangle_0.  \label{eq:3.3}
\end{eqnarray}
The term $\langle H_0\rangle_0$ is just giving the ground state energy
of the free system. It is the sum of the excitation energies of the atoms and the
black-body radiation energy. 
The later is $0$ for $T\to 0$, since we use the normal ordering in the definition
of the free photon Hamiltonian (\ref{eq:2.3}). 
The normal ordering is necessary for the Hamiltonian to be well-defined.
The $\langle H_0\rangle_0$-term will be ignored in the following. 
The expectation value of the $S$-matrix 
is given by perturbation theory.
\begin{equation} 
  \langle S\rangle_0=1-\sum_{n=1}^\infty(-1)^n
  \!\int_0^\beta \! d\tau_1\dots d\tau_n
  \langle T_\tau H_1(\tau_1)\dots H_1(\tau_n)\rangle_0 \label{eq:3.4}
\end{equation}
where $H_1(\tau)$ is the Heisenberg representation for the 
interaction Hamiltonian $H_1$ with imaginary time $\tau=-it$. It is well known
that in $\ln\langle S\rangle_0$ only connected graphs contribute \cite{abrikosov,mahan}.
We have 
\begin{eqnarray} 
  \ln\langle S\rangle_0&=&
  \sum_{m=-\infty}^\infty\sum_{n=2}^\infty\frac{1}{2n}\sum_{\vec x_1\dots\vec x_n}
  \hat\alpha_0(i\omega_m,\vec x_1)\hat D^0_{i_1i_2}(i\omega_m,\vec x_1-\vec x_2) 
  \nonumber\\
  & & \qquad \dots \hat\alpha_0(i\omega_m,\vec x_n)
      \hat D^0_{i_ni_1}(i\omega_m,\vec x_n-\vec x_1) \label{eq:3.5}
\end{eqnarray}
where the last sum goes over the positions $\vec x_i$ of the dipoles with 
$\vec x_1 \neq\vec x_2 \cdots$ for not including the divergent self-energy 
contributions. $\alpha_0$ is the atomic polarization given by
\begin{equation} 
  \alpha_{ij}(\tau,\vec x-\vec x\,')
  = -\langle T_\tau \,q_i(\tau,\vec x)\,q_j(0,\vec x\,')\rangle_0
  = \alpha_0(\tau,\vec x)\delta_{ij}\delta_{xx'} \label{eq:3.6}
\end{equation}
for isotropic atoms. Since $\alpha_0$ obeys the KMS periodicity
$\alpha_0(\tau)=\alpha_0(\tau+\beta)$, it can be expand into a 
Fourier series
\begin{equation} 
  \alpha_0(\tau,\vec x)=\frac{1}{\beta}\sum_{m=-\infty}^\infty 
  \hat \alpha_0(i\omega_m,\vec x)\,e^{-i\omega_m\tau}, \label{eq:3.7}
\end{equation}
where 
\begin{equation} 
  \omega_m=\frac{2\pi m}{\beta}\label{eq:3.8}
\end{equation}
are the Matsubara frequencies. The Fourier coefficients in (\ref{eq:3.7}) 
are given by
\begin{equation}
  \hat\alpha_0(w,\vec x)=\int_0^\beta d\tau 
  \alpha_0(\tau,\vec x) e^{w\tau}, \label{eq:3.9}
\end{equation}
using the short-hand notation
\begin{equation}
  w=i\omega_m=i\frac{2\pi m}{\beta}. \label{eq:3.10}
\end{equation}
The Green's function $\hat D^0_{ij}(\omega,\vec x- \vec y)$ in 
(\ref{eq:3.5}) is the Fourier transformed free Green's function of the total
electric field given by 
\begin{equation}
  \label{eq:3.11}
   D^0_{ij}(\tau,\vec x- \vec y)
   =\partial_\tau^2
     \langle T_\tau\,A_i(\tau,\vec x)\,A_j(0,\vec y)\rangle_0 +
     \Phi_{ij}(\vec x-\vec y)
\end{equation}
In momentum space we get
\begin{eqnarray}
  \hat D^0_{ij}(\omega,\vec k)
  &=&\frac{\omega^2(\delta_{ij}-\displaystyle\frac{k_ik_j}{k^2})}
   {k^2-\omega^2}-\frac{k_ik_j}{k^2} 
  = \frac{\omega^2\delta_{ij}-k_ik_j}{k^2-\omega^2}\nonumber\\
  &=& -\delta_{ij}+\frac{k^2}{k^2-\omega^2}\,
      \sum_{\lambda=1,2}\xi^\lambda_i(\vec k)\xi^\lambda_j(\vec k)
  \label{eq:3.12}
\end{eqnarray}
We get the energy in (\ref{eq:3.3}) by differentiating (\ref{eq:3.5}) with
respect to $\beta$ and using $\partial_\beta\omega_m=-\omega_m/\beta$. Hence
\begin{eqnarray}
  E&=&\frac{1}{\beta}\sum_{m=-\infty}^\infty
      \omega_m\partial_{\omega_m}
      \sum_{n=2}^\infty\frac{1}{2n}\sum_{\vec x_1\cdots\vec x_n}
      \hat\alpha_0(i\omega_m,\vec x_1)
      \hat D^0_{i_1i_2}(i\omega_m,\vec x_1-\vec x_2)\dots
      \nonumber\\
   & &\qquad\qquad 
      \dots\,\hat\alpha_0(i\omega_m,\vec x_n)
      \hat D^0_{i_ni_1}(i\omega_m,\vec x_n-\vec x_1). \label{eq:3.13}
\end{eqnarray}
To calculate the Casimir force with finite temperature we need the free energy
given by 
\begin{eqnarray}
  F&=&-\frac{1}{\beta}\ln\left<S\right>_0 \nonumber\\
   &=&-\frac{1}{\beta}\sum_{m=-\infty}^\infty
      \sum_{n=2}^\infty\frac{1}{2n}\sum_{\vec x_1\cdots\vec x_n}
      \hat\alpha_0(i\omega_m,\vec x_1)
      \hat D^0_{i_1i_2}(i\omega_m,\vec x_1-\vec x_2)\dots
      \nonumber\\
   & &\qquad\qquad 
      \dots\,\hat\alpha_0(i\omega_m,\vec x_n)
      \hat D^0_{i_ni_1}(i\omega_m,\vec x_n-\vec x_1). \label{eq:3.13.1}
\end{eqnarray}
In the limit $T \to 0\,(\beta\to\infty)$ the two energies $E$
and $F$ are the same. Indeed the summation over 
the frequencies $\omega_m$ is replaced by an integral:
\begin{equation}
  \label{eq:3.14}
  \frac{1}{\beta}\sum_{m=-\infty}^\infty f(\omega_m) \longrightarrow 
  \frac{1}{2\pi}\int_{-\infty}^\infty du\,f(u)
  \quad\text{for}\quad T\to 0\,,
\end{equation}
and $E$ then transforms into $F$ by partial integration. 

It is convenient to put the position dependence of the dipoles in the 
atomic polarization function: 
$\alpha_0(\vec x)=\sum_{\vec R} \alpha_0\delta(\vec x- \vec R)$ where $\vec R$
runs over all positions of the dipoles. For not including
self-energy contributions, we have to replace $D^0$ by $D_0'$: 
\begin{equation}
  \label{eq:b1}
  D'_0(iw_n,\vec x)=\left\{ 
  \begin{array}{ccc}
    D^0(iw_n,\vec x) &\text{for}& \vec x\neq \vec 0\\
    0 &\text{for}& \vec x = \vec 0
  \end{array}
  \right.\,.
\end{equation}
The integration over $x$ can be considered as matrix multiplications.
Then the spectral densities of the energies
(\ref{eq:3.13},\ref{eq:3.13.1}) can be
written as  
\begin{equation}
  \label{eq:3.15}
  E(\omega)=\frac{1}{2}w\,\partial_w\sum_{n=2}^\infty \frac{1}{n}
    \tr \bigl( (\alpha_0 D'_0)^n \bigr)
\end{equation}
\begin{equation}
  \label{eq:3.16}
  F(w)=-\frac{1}{2}\sum_{n=2}^\infty \frac{1}{n}
    \tr \left( (\alpha_0 D'_0)^n \right)
\end{equation}
The two energy densities are finite, since self-energy
contributions ($\vec x_i=\vec x_{i+1}$) are excluded by (\ref{eq:b1}).

Next we consider the limit to a continuous distribution of the dipoles. 
The sums over the positions are then replaced by integrals over the 
whole space. The polarization function is now
\begin{equation}
  \label{eq:b3}
  \alpha_0(iw_n,\vec x)=\alpha_0(iw_n)\theta_K(\vec x)\,,
\end{equation}
where $\theta_K(\vec x)=1$ for $\vec x$ in the region $K$ occupied by the
dielectric and $0$ otherwise. The polarization $\alpha_0(iw_n)$ is replaced by
its density.
$D_0'$ is an integral operator with kernel
$D_0'(i\omega_n,\vec x-\vec y)$. 
The limit only exists if $D'_0$ in (\ref{eq:3.15}, \ref{eq:3.16})
contains an ultraviolet cut-off in momentum space or a cut-off at 
small distances in $x$-space. This can be seen as follows.
According to (\ref{eq:3.15}) $(\alpha_0D'_0)^2$ must be a trace-class
operator. This is true if $\alpha_0 D'_0$ is a Hilbert - Schmidt
operator, that means its integral kernel must be square integrable
\begin{equation}
  \label{eq:31}
  \int d^3y\int d^3x\vert\alpha_0(\vec x)D'_0(\vec x-\vec y)\vert^2
  =\Vert\alpha_0D'_0\Vert^2_{\rm HS}
  =\alpha_0^2\cdot {\rm vol}(K)\Vert D'_0
  \Vert_2^2,<\infty.
\end{equation}
Note that the square of a Hilbert - Schmidt operator is trace-class,
and the same is true for all higher powers in (\ref{eq:3.15}, 
\ref{eq:3.16}). The dielectric must have a finite volume and the Green's function 
be square integrable. But a glance to (24) shows that an ultraviolet cut-off
is necessary for this. The problem comes from the
dipole - dipole interaction. 

In $x$-space the dipole - dipole term $\Phi_{ij}(\vec x)$ is not
square integrable at $\vec x=0$. Therefore, we introduce a spatial
cut-off at a minimum average
distance $a$ between the atoms as it was also done
by Marachevsky \cite{marachevsky}. We correct the Green's function $D^0$
for small distances as follows 
\begin{equation}
  \hat D'_0(\omega,\vec x)=\hat D^0(\omega, \vec x)\theta (x-a)
  = \hat D^0(\omega,\vec x) -\hat G^0(\omega,\vec x), \label{eq:3.17}
\end{equation}
where
\begin{equation}
  \hat G^0(\omega,\vec x)=\hat D^0(\omega,\vec x)\theta(a-x). \label{eq:3.18}
\end{equation}

We show next, that the splitting of $D'_0=D^0-G^0$ leads to corresponding
splitting in the energy $E=E_D-E_G$. 
In the sums (\ref{eq:3.15},\ref{eq:3.16}) we introduce the first order term
in $\alpha_0$, which is $0$ since $\alpha_0D'_0(\vec 0)=0$, and get 
\begin{eqnarray}
  \label{eq:3.19}
  -\sum_{n=1}^\infty \frac{1}{n}\tr \bigl( (\alpha_0 D'_0)^n \bigr)
  &=&\tr \bigl(\ln (1-\alpha_0 D^0 + \alpha_0 G^0)\bigr) \nonumber\\
  &=& \tr \Bigl( \ln \bigl((1+\alpha_0 G^0)(1-(1+\alpha_0 G^0)^{-1} 
                            \alpha_0 D^0)\bigr)
         \Bigr)\nonumber\\
  &=& \tr \bigl( \ln (1+\alpha_0 G^0)+ \ln (1 -\alpha D^0)\,
\bigl),
\end{eqnarray}
where $\alpha$ is the macroscopic polarizability given by
\begin{equation}
  \label{eq:3.20}
  \alpha=(1+\alpha_0 G^0)^{-1}\alpha_0 .
\end{equation}
If we express $\alpha_0$ by $\alpha$ in the first term of (\ref{eq:3.19}) we get
$1+\alpha^0G^0=(1-\alpha G^0)^{-1}$.
We would like to split the trace in (\ref{eq:3.19}) into two separate
traces. As discussed above this is only possible with an ultraviolet
cut-off $e^{-\delta k}$ in the Green's functions.
For the inner energy $E$ we then get
\begin{equation}
  \label{eq:3.21}
  E=\lim_{\delta\to 0}(E^\delta_D-E^\delta_G)
\end{equation}
with
\begin{equation}
  \label{eq:3.22}
  E^\delta_D(w)=w\partial_w\frac{1}{2} \sum_{n=1}^\infty \frac{1}{n}
  \tr \left( (\alpha D_\delta^0)^n \right)
\end{equation}
and
\begin{equation}
  \label{eq:3.23}
  E^\delta_G(w)=w\partial_w\frac{1}{2} \sum_{n=1}^\infty \frac{1}{n}
    \tr \left( (\alpha G_\delta^0)^n \right)
\end{equation}
and similarly for the free energy $F$.
In the limit $\delta\to 0$ $E$ remains finite but $E^\delta_D$ and 
$E^\delta_G$ are both
divergent. In the limit $a\to 0$ also $E$ is divergent. 
The energy $E^\delta_D$ has the similar form as in (\ref{eq:3.15}), 
only that the atomic polarizability is replaced by the macroscopic
polarizability $\alpha$ defined in (\ref{eq:3.20}) and $D'_0$ is replaced by
the cut-off Green's function $D^0_\delta$. 
As the cut-off parameters $a$ and $\delta$ go to $0$, we have
\begin{equation}
  \label{eq:3.23.1}
  G_\delta^0(\omega, \vec x) \to -\frac{1}{3}\delta(\vec x) 
\end{equation}
and we get from (\ref{eq:3.20})
\begin{equation}
  \label{eq:3.24}
  \alpha(\omega,\vec x,\vec x\,')=
  \frac{\alpha_0(\omega,\vec x)}{1-\frac{1}{3}\alpha_0(\omega,\vec x)}
  \delta(\vec x-\vec x\,')
\end{equation}
the Lorentz-Lorenz relation for the macroscopic polarizability. We shall use
this relation in later calculations.

$E^\delta_D$ is for $\delta,a\to 0$  formally equivalent to the energy $E_D$ 
that is usually calculated in the
Casimir problem.  
To bring the energy $E_D^\delta$ in a convenient form, we perform the
derivative in (38) using the relation 
\begin{equation}
  \label{eq:3.26}
  w\partial_wD^0_\delta=2\,\bigl(D^0_\delta+(D^0_\delta)^2\bigr)\,.
\end{equation}
(\ref{eq:3.22}) then becomes
\begin{equation}
  \label{eq:3.24.1}
  E_D^\delta(w)=\frac{1}{2}{\rm Tr}{\sum_n}'
         \bigl((w\partial_w \alpha)+2(1+\alpha)\bigr)
         D^0_\delta \left(\alpha D^0_\delta\right)^n \,,
\end{equation}
where the prime in the summation means that we just sum up in orders of
$\alpha>0$. The last part $\sum_n D^0(\alpha D^0)^n$ is 
the Green's function of the total electromagnetic field
\begin{eqnarray}
  \label{eq:3.27}
  \lefteqn{\langle E_i(\vec x),E_j(\vec x\,')\rangle 
          = D^0(\vec x-\vec x\,')}\nonumber\\
  & & \quad{}+\sum_{n=1}^\infty \int dy_1\dots dy_{n}
      D^0_{ik_1}(\vec x-\vec y_1)\alpha(y_1) \dots
      \alpha(y_{n})D^0_{k_nj}(\vec y_n-\vec x\,') \,,
\end{eqnarray}
where $\alpha$ is given by (\ref{eq:3.24}). 
One can get this Green's function also by ordinary perturbation theory and 
Dyson resummation. It will be discussed extensively in the next section.
Note that in (44) no cut-off  is needed.\\
With $\vec B=i/w\,\vec\nabla\wedge\vec E$ and the relation 
$\langle\eps\vec E(\vec x)^2\rangle=\langle\vec B(\vec x)^2\rangle+ div\,$  we
finally 
get for energy density 
\begin{equation}
  \label{eq:3.28}
  e_D(w,\vec x)=  \frac{1}{2}\Bigl( 
      \bigl<\bigl(\partial_ww\,\eps(w,\vec x)\bigr)
      E_i(\vec x)E_i(\vec x)\bigr>'
      +\bigl<B_i(\vec x)B_i(\vec x )\bigr>'\Bigr)
 \end{equation}
where $\eps=1+\alpha$ and the prime means, that the $0$-th order in $\alpha$
is suppressed.
This corresponds only to the $T_{00}$-component of electromagnetic stress tensor if
there is no dispersion. 
The energy density in (\ref{eq:3.28}) was also used by Candelas
\cite{candelas} (cited by \cite{kupiszewska}). 
However this energy density is only well defined outside the dielectric. The
total energy is infinite. 

For two disconnected dielectrics it is possible to express the energy $F$ by
three parts: two containing just the energies of the isolated dielectrics and
one containing the interaction energies of the to bodies. 
Let $\alpha_0(\vec x)=\alpha_0^A(\vec x) +\alpha_0^B(\vec x)$,
where $A$ and $B$ are the regions occupied by the two dielectrics. 
We then have
\begin{eqnarray}
  \label{eq:b4}
  F(w)&=&\frac{1}{2}\tr\Bigl(\ln\bigl(1-(\alpha_0^A+\alpha_0^B)D'_0\bigr)\Bigr)
      \nonumber\\
      &=&\frac{1}{2}\tr\Bigl(\ln\bigl(1-\alpha_0^AD'_0\bigr)\Bigr)
        +\frac{1}{2}\tr\Bigl(\ln
         \bigl(1-(1-\alpha_0^AD'_0)^{-1}\alpha_0^BD'_0\bigr)\Bigr)
\end{eqnarray}
The first term is just the energy 
$F_A(w)=\frac{1}{2}\tr\bigl(\ln(1-\alpha_0^AD'_0)\bigr)$ 
of the dielectric in the region $A$. For the
second term we use 
$(1-\alpha_0^AD'_0)^{-1}=1+(1-\alpha_0^AD'_0)^{-1}\alpha_0^AD'_0$ and we get
\begin{equation}
  \label{eq:b5}
  F(w)=F_A(w)+F_B(w) 
       +\frac{1}{2}\tr\Bigl(\ln
         \bigl(1-(1-\alpha_0^AD'_0)^{-1}\alpha_0^AD'_0
                 (1-\alpha_0^BD'_0)^{-1}\alpha_0^BD'_0\bigr)\Bigr)
\end{equation}
where
\begin{equation}
  \label{eq:b5.1} 
  F_B(w)=\frac{1}{2}\tr\Bigl(\ln\bigl(1-\alpha_0^BD'_0\bigr)\Bigr).
\end{equation}
For the last expression in (\ref{eq:b5}) we use
\begin{equation}
  \label{eq:b6}
  (1-\alpha_0^A(D^0-G^0))^{-1}\alpha_0^A=(1-\alpha_AD^0)^{-1}\alpha_A\,,
\end{equation}
with $\alpha_A=(1-\alpha_0^AG^0)^{-1}\alpha_0^A$.
Since $\alpha_AD'_0\alpha_B=\alpha_AD^0\alpha_B$ for separated regions $A$ and
$B$, we get for the last term in (\ref{eq:b5})
\begin{equation}
  \label{eq:b7}
  F_{AB}(w)=\frac{1}{2}\tr\Bigl(\ln
            \bigl(1-\alpha_AD_A\alpha_BD_B\bigr)\Bigr)\,,
\end{equation}
where
\begin{equation}
  \label{eq:b8}
  D_A=(1-D^0\alpha_A)^{-1}D^0
\end{equation}
and similar for $D_B$. $D_A$ is just the Green's function of the
electromagnetic field for only one dielectric in the region $A$ as shall be
shown in the next section. 
Hence we have for the free energy
\begin{equation}
  \label{eq:b9}
  F(w)=F_A(w)+F_B(w)+F_{AB}(w)
\end{equation}
All the three energies are finite. For the energies $F_A$ and $F_B$ one has to
take the finite distance between the atoms into account. $F_{AB}$ is the only term
where the distance between the dielectrics enters. Only this energy
contributes to the Casimir force. It remains also finite in the limit $a\to
0$. Then for the polarizations $\alpha_A$ and $\alpha_B$ the Lorentz-Lorenz
formula (\ref{eq:3.24}) holds.     

To calculate the force between two dielectrics we will use the stress tensor
of the electromagnetic field. The force is given by a contour integral over a
region containing only one of the two bodies. Only the stress outside the
dielectrics contribute to the force. It can be calculated as usual from 
the expression:
\begin{eqnarray}
  \label{eq:3.30}
  T_{ik}(\vec x)&=&\frac{1}{2}\delta_{ik}\left<E_l(\vec x) \,E_l(\vec x) \right>'
    -\left<E_i(\vec x) \,E_k(\vec x) \right>'\nonumber\\
  && +\frac{1}{2}\delta_{ik}\left<B_l(\vec x) \,B_l(\vec x) \right>'
     -\left<B_i(\vec x) \,B_k(\vec x) \right>'
\end{eqnarray}
The prime again means that the $0$-th order in $\alpha$ is suppressed.


\section{\label{sec:basic}Basic integral equation}


To calculate the Casimir force from (53) one has to calculate the 
Green's function of the total electromagnetic field
$D_{ij}(\vec x,\vec x\,')=\left<E_i(\vec x)\,E_j(\vec x\,')\right>$ given in
(\ref{eq:3.27}). Let $\alpha(w,\vec x)=\alpha(w)\,\theta_K(\vec x)$, where $K$ is
the region occupied by the dielectric. The Green's function can be expressed 
as an integral equation:
\begin{equation}
  \label{eq:4.3}
  D_{ij}(\omega,\vec x,\vec x\,')=D^0_{ij}(\omega,\vec x-\vec x\,')
  +\,\alpha(\omega)\int_K d^3y D^0_{ik}(\omega,\vec x-\vec y)
    D_{kj}(\omega,\vec y,\vec x\,'),
\end{equation}
where no cut-off is necessary.
This is our basic integral equation. It separates the many-body problem from
the geometry. The latter only enters in the region of integration $K$
occupied by the dielectric. For simplicity we have assumed one
kind of dielectrics and vacuum outside $K$, the generalization to 
more dielectrics is straightforward. 

From the integral equation
one can derive a differential equation for $D$. 
From 
\begin{equation}
  \label{eq:4.4}
  \bigl((\omega^2+\lap )\delta_{ik}-\partial_i\partial_k\bigr)D^0_{kj}(\omega,\vec x)=-\omega^2
  \delta_{ij}(\vec x)
\end{equation}
one easily derives
\begin{equation}
  \label{eq:4.5}
  \left(\left(\eps(\omega,\vec x)\omega^2+\lap\right)\delta_{ik}-\partial_i\partial_k\right)
  D_{kj}(\omega,\vec x,\vec x\,')
  =-\omega^2 \delta_{ij}(\vec x-\vec x\,')
\end{equation}
The boundary conditions can also be derived from the integral equation
(\ref{eq:4.3}).
In $D^0$ only the part of the dipole-dipole interaction is responsible for
non-continuous boundary conditions. Hence the dipole-dipole approximation is
necessary to produce hard boundary conditions and in the same time it is responsible
for divergences for connected bodies, if the self-energy contributions are ignored. 

A solution of the Casimir problem by solving the differential equation with
boundary conditions was
given by Lifshitz \cite{lifshits} for simple plain geometry. 
The solution of the boundary-value problem is much harder than the solution 
of (\ref{eq:4.3}). In principle the linear equation (\ref{eq:4.3}) can be solved
numerically for any geometry. In the following we are interested in
cases where this solution can be obtained analytically. We solve directly
the integral equation without the use of any boundary condition.\\
In the shorthand notation the basic integral equation writes as follows:
\begin{equation}
  \label{eq:4.6}
  D=D^0+\alpha D^0\theta_KD
\end{equation}
For a infinite dielectric ($K=\bbbr^3$) the equation can be solved in
$k$-space:
\begin{equation}
  \label{eq:4.7}
  D^1=D^0+\alpha D^0D^1=(1-\alpha D^0)^{-1}D^0
\end{equation}
leading to 
\begin{equation}
  \label{eq:4.8}
  D^1_{ij}(\omega,\vec k)=\frac{1}{\eps}\,
  \frac{\eps\omega^2\delta_{ij}-k_ik_j}{k^2-\eps\omega^2}
  =\frac{1}{\eps}\,D^0_{ij}(\sqrt{\eps}\omega,\vec k)
\end{equation}
Let
\begin{equation}
  \label{eq:4.9}
  \theta_{\bar K}=1-\theta_K
\end{equation}
Substitute $\theta_K$ in the basic integral equation (\ref{eq:4.6}) gives
\begin{equation}
  \label{eq:4.10}
  D=D^0+\alpha D^0D-\alpha D^0\theta_{\bar{K}}D .
\end{equation}
Hence
\begin{equation}
  \label{eq:4.11}
  (1-\alpha D_0)D=D^0-\alpha D^0\theta_{\bar K}D
\end{equation}
And using (\ref{eq:4.7}) we get
\begin{equation}
  \label{eq:4.12}
  D=D^1-\alpha D^1\theta_{\bar K}D
\end{equation}
the complementary basic equation.
To obtain the Casimir force we have to compute
\begin{equation}
  \label{eq:4.13}
  \theta_KD\theta_{\bar K}=\theta_KD^1\theta_{\bar K}-\alpha
  \theta_KD^1\theta_{\bar K}D\theta_{\bar K}.
\end{equation}
Here we substitute (\ref{eq:4.6})
\begin{equation}
  \label{eq:4.14}
  \theta_K D\theta_{\bar K}
   =\theta_KD^1\theta_{\bar K}-\alpha\theta_KD^1\theta_{\bar K}D^0\theta_{\bar K}
   -\alpha^2(\theta_KD^1\theta_{\bar K}D^0)\theta_KD\theta_{\bar K}.
\end{equation}
Now the same quantity as on the l.h.s has appeared, which allows the
solution 
\begin{equation}
  \label{eq:4.15}
  \theta_KD\theta_{\bar K}=(1+\alpha^2\theta_KD^1\theta_{\bar K}D^0)^{-1}
  \theta_KD^1\theta_{\bar K}(1-\alpha D^0\theta_{\bar K}).
\end{equation}
The above formal solution (\ref{eq:4.15}) is very useful in the case where the
operator in the inverse operates by simple multiplication:
\begin{equation}
  \label{eq:4.16}
  (\theta_KD^1\theta_{\bar K}D^0)\theta_KD^1\theta_{\bar K}=\gamma(w)
  \theta_KD^1\theta_{\bar K}.
\end{equation}
As we will see this is indeed true for various simple geometries. Then we
obtain from (\ref{eq:4.15}) and (\ref{eq:4.6}) the solutions
\begin{equation}
  \label{eq:4.17}
  \theta_KD\theta_{\bar K}=\frac{1}{1+\alpha^2\gamma}\theta_KD^1\theta_{\bar K}
  (1-\alpha D^0\theta_{\bar K}),
\end{equation}
\begin{equation}
  \label{eq:4.18}
  \theta_{\bar K}D\theta_{\bar K}=\theta_{\bar K}D^0\theta_{\bar K}
  +\frac{\alpha}{1+\alpha^2\gamma}
  \theta_{\bar K}D^0\theta_KD^1\theta_{\bar K}(1-\alpha D^0\theta_{\bar K}).
\end{equation}


\section{\label{sec:plain}Plain geometries}

Let us first consider an infinitely thick plate in the region $-\infty<
x<0$. In this case we choose the polarization vectors (\ref{eq:2.10}) as
follows: Let $\vec n=(1,0,0)$, then
\begin{equation}
  \vec \xi_1(\vec k)=\frac{\vec k\wedge\vec n}{|\vec k\wedge\vec n|}
  =\frac{1}{p}(0,-k_3,k_2)\label{eq:5.1}
\end{equation}
\begin{equation}
  \vec \xi_2(\vec k)=\frac{\vec k\wedge(\vec k\wedge\vec n)}
                          {|\vec k\wedge(\vec k\wedge\vec n)|}
  =\frac{1}{pk}(p^2,-k_2k_1,-k_3k_1),\label{eq:5.2}
\end{equation}
where $p^2=k_2^2+k_3^2$ and $k^2=k_1^2+p^2$. $D^0$ (\ref{eq:3.14}) can be decomposed
accordingly 
\begin{equation}
D^0=D^0_1+D^0_2,\label{eq:5.3}
\end{equation}
\begin{equation}
  D^0_{1ij}(x)=\frac{1}{2\pi}\int dk_1\,
  \xi_{1i}\xi_{1j}\frac{w^2}{k_1^2-s_0^2}e^{ik_1x},\label{eq:5.4}
\end{equation}
where $s_0^2=w^2-p^2, w=i\omega_n$, and
\begin{equation}
  D^0_{2ij}(x)=\delta(x)(\xi_{1i}\xi_{1j}-\delta_{ij})
  +\frac{1}{2\pi} 
  \!\!\int\!\! dk_1\,\xi_{2i}\xi_{2j}\frac{k^2}{k_1^2-s_0^2}e^{ik_1x}.\label{eq:5.5}
\end{equation}
The integrals can be easily computed:
\begin{equation}
  D^0_{1ij}(x)=\xi_{1i}\xi_{1j}\frac{iw^2}{2s_0}\Bigl(\theta(x)e^{is_0x}
  +\theta(-x)e^{-is_0x}\Bigl)\label{eq:5.6}
\end{equation}
\begin{equation}
  D^0_{2ij}(x)=\delta(x)(\xi_{1i}\xi_{1j}-\delta_{ij})
  + \xi_{2i}\xi_{2j}\frac{iw^2}{2s_0}
    \left(\theta(x)e^{is_0x}+\theta(-x)e^{-is_0x}\right),\label{eq:5.7}
\end{equation}
where $\vec\xi_2=1/wp\,(p^2,i\partial_xk_2,i\partial_xk_3)$ due to the Fourier
transformation. The corresponding formulae for the homogeneous
dielectric (\ref{eq:4.8}) are similar: 
\begin{equation}
  \label{eq:5.8.0}
  D^1_{1ij}(x)=\xi_{1i}\xi_{1j}\frac{iw^2}{2s_1}\Bigl(\theta(x)e^{is_1x}
  +\theta(-x)e^{-is_1x}\Bigl)
\end{equation}
and
\begin{equation}
  D^1_{2ij}(x)=
  \frac{1}{\eps}\delta(x)(\xi_{1i}\xi_{1j}-\delta_{ij})
  + \xi'_{2i}\xi'_{2j}\frac{i\eps w^2}{2s_1}
  \left(\theta(x) e^{is_1x}+\theta(-x)e^{-is_1x}\right),\label{eq:5.8}
\end{equation}
where
\begin{equation}
  s_1^2=\eps w^2-p^2,\quad\vec \xi_2\,'
       =\frac{1}{\sqrt{\eps}wp}(p^2,i\partial_xk_2,i\partial_xk_3).
\end{equation}
The above decomposition into the two transverse modes (TE and TM mode) of the radiation
field leads to a similar decomposition in (\ref{eq:4.17}) and
(\ref{eq:4.18}). There is no 
mode mixing. Details of the calculation are given in Appendix \ref{sec:app1}. For
(\ref{eq:4.17}) we obtain
\begin{equation}
  \theta(-x)D(x,x')\theta(x')=
  \Bigl(\frac{iw^2}{s_1+s_0}\xi_{1i}\xi_{1j}^*
  +\frac{i\sqrt{\eps}w^4}{(s_1+s_0)(p^2+s_1s_0)}
  \xi_{2i}\xi_{2j}^*\Bigl)
  \theta(-x)e^{-is_1x}\theta(x')e^{is_0x'},\label{eq:5.9}
\end{equation}
and for the reflection (\ref{eq:4.18})
\begin{eqnarray}
  \lefteqn{\theta(x)D(x,x')\theta(x')
           =\theta(x)D^0(x,x')\theta(x')}\nonumber\\
  & &+\frac{iw^2}{2s_0}\Bigl(r_1(w,p)\,\xi_{1i}\xi_{1j}^*
     +r_2(w,p)\,\xi_{2i}\xi_{2j}^*\Bigl)
     \theta(x)e^{is_0x}\theta(x')e^{is_0x'}\,\label{eq:5.10}
\end{eqnarray}
where the second term herein represents the reflection at the boundary $x=0$
with the reflection factors $r_1$ and $r_2$ for the TE and the TM mode given by
\begin{equation}
  \label{eq:5.10.1}
  r_1(w,p)=-\frac{s_1-s_0}{s_1+s_0}
  =-\frac{\sqrt{\eps(w)w^2-p^2}-\sqrt{w^2-p^2}}
        {\sqrt{\eps(w)w^2-p^2}+\sqrt{w^2-p^2}}
\end{equation}
and
\begin{equation}
  \label{eq:5.10.2}
  r_2(w,p)=-\frac{s_1-\eps s_0}{s_1+\eps s_0}
  =-\frac{\sqrt{\eps(w)w^2-p^2}-\eps(w)\sqrt{w^2-p^2}}
        {\sqrt{\eps(w)w^2-p^2}+\eps(w)\sqrt{w^2-p^2}}\,.
\end{equation}
For later use we rename the Green's function for the region
$A: x<0$ by $D^A(x,x')$.
Now we fill also the region $B: x>a>0$ with the dielectric with the same
$\eps$, in between $0<x<a$ we have vacuum. The Green's functions for the
region $B$ alone are obtained from the above formulae by substituting $x$
by $a-x$:
\begin{equation}
  \theta_BD^B\theta_{\bar B} =
  \Bigl(\frac{iw^2}{s_1+s_0}\xi_{1i}\xi_{1j}^*
       +\frac{i\sqrt{\eps}w^4}{(s_1+s_0)(p^2+s_1s_0)}\xi_{2i}\xi_{2j}^*
  \Bigl)
  \theta(x-a)e^{-is_1(a-x)}\theta(a-x')e^{is_0(a-x')},\label{eq:5.14}
\end{equation}
\begin{eqnarray}
  \lefteqn{\theta_{\bar B}D(x,x')\theta_{\bar B}
           =\theta(a-x)D^0(x,x')\theta(a-x')}\nonumber\\
  & & {}+\frac{iw^2}{2s_0}
  \Bigl(r_1(w,p)\xi_{1i}\xi_{1j}^*
         +r_2(w,p)\xi_{2i}\xi_{2j}^*
  \Bigl)
  \theta(a-x)e^{-is_0x}\theta(a-x')e^{-is_0x'}e^{2is_0a}.
  \nonumber\\&&
  \label{eq:5.15}
\end{eqnarray}
According to (\ref{eq:4.6}) the total Green's function is the solution of the integral 
equation
\begin{equation}
D=D^0+\alpha D^0\theta_AD+\alpha D^0\theta_BD.\label{eq:5.16}
\end{equation}
Inserting the definitions of $D^A$ and $D^B$, we get
\begin{equation}
D=D^A+\alpha D^A\theta_BD\label{eq:5.17}
\end{equation}
\begin{equation}
D=D^B+\alpha D^B\theta_AD.\label{eq:5.18}
\end{equation}
Let us introduce the characteristic function
\begin{equation}
\theta_C=1-\theta_A-\theta_B\label{eq:5.19}
\end{equation}
for the region between the two dielectrics.
Then by combining the equations we find for the polarization $\lambda$:
\begin{eqnarray}
  \theta_AD_\lambda\theta_C
  &=&\theta_AD^A_\lambda\theta_C+\alpha\theta_A
      D^A_\lambda\theta_BD^B_\lambda\theta_C
      +\alpha^2\theta_A
      D^A_\lambda\theta_BD^B_\lambda\theta_AD_\lambda\theta_C \nonumber\\
  &=& \frac{1}{1-\alpha^2\gamma_\lambda}(\theta_AD^A_\lambda\theta_C+\alpha
       \theta_AD^A_\lambda\theta_BD^B_\lambda\theta_C).\label{eq:5.20}
\end{eqnarray}
Here we have used the fact that the operator
\begin{equation}
(\theta_AD^A_\lambda\theta_BD^B_\lambda)\theta_AD^A_\lambda\theta_B=
\gamma_\lambda\theta_AD^A_\lambda\theta_B\label{eq:5.21}
\end{equation}
operates simply by multiplication, this is shown in Appendix \ref{sec:app1}. 
Substitution into (\ref{eq:5.18}) gives
\begin{equation}
  \theta_CD_\lambda\theta_C=\theta_CD^B_\lambda\theta_C
  +\mu_\lambda \alpha
  \left(\theta_CD^B_\lambda\theta_AD^A_\lambda\theta_C
        +\alpha\theta_CD^B_\lambda\theta_AD^A_\lambda
        \theta_BD^B_\lambda\theta_C
  \right)\,,\label{eq:5.21.1}
\end{equation}
where
\begin{equation}
\mu_\lambda=\frac{1}{1-\alpha^2\gamma_\lambda}.\label{eq:5.22}
\end{equation}
Substitution of $D^B_\lambda$ by its defining equation
$D^B_\lambda=D^0_\lambda+\alpha D^0_\lambda\theta_BD^B_\lambda$ leads to 
\begin{eqnarray}
  \label{eq:5.22.0}
  \theta_CD_\lambda\theta_C&=&\theta_CD_\lambda^0\theta_C+
  \mu_\lambda\alpha\theta_CD_\lambda^0\theta_AD_\lambda^A\theta_C+
  \mu_\lambda\alpha\theta_CD_\lambda^0\theta_BD_\lambda^B\theta_C\nonumber\\
  &&{}+\mu_\lambda\alpha^2
    \theta_CD_\lambda^0\theta_AD_\lambda^A\theta_BD_\lambda^B\theta_C+
  \mu_\lambda\alpha^2
  \theta_CD_\lambda^0\theta_BD_\lambda^B\theta_AD_\lambda^A\theta_C 
\end{eqnarray} 
The calculation of (\ref{eq:5.22.0}) is given in Appendix \ref{sec:app1} with
the following results: 
\begin{eqnarray}
  D_\lambda(x,x')
  &=& D_\lambda^0(x,x')\nonumber\\
  & & {}+ \mu_\lambda\,r_\lambda\,\frac{iw^2}{2s_0}\,
     \xi_\lambda\xi_\lambda^*\Bigl(e^{is_0(x+x')}+e^{-is_0(x+x'-2a)}\Bigl)
  \nonumber\\
  & & {}+\mu_\lambda\, r_\lambda^2\,e^{2is_0a}\,
  \frac{iw^2}{2s_0}\,\xi_\lambda\xi_\lambda^*\Bigl( e^{is_0(x-x')}+e^{-is_0(x-x')}
                                        \Bigl),
  \label{eq:5.22.1}
\end{eqnarray}
\begin{equation}
  \mu_\lambda=\frac{1}{1-r_\lambda^2\, e^{2is_0a}},\label{eq:5.23}
\end{equation}
with $\lambda=1,2$ for the two modes and where $r_\lambda$ are the reflection
factors given in (\ref{eq:5.10.1}) and 
(\ref{eq:5.10.2}). 
Since the 0-th order does not contribute later on, we introduce
$D^W=D-D^0$:
\begin{equation}
  D^W_{ij}(w;x,x')
  =\xi_{1i}(x)\xi_{1j}(x')^*f^1(w;x,x')
  +\xi_{2i}(x)\xi_{2j}(x')^*f^2(w;x,x'),\label{eq:5.26}
\end{equation}
where
\begin{equation}
  \vec \xi_1(\vec k)=\frac{1}{p}(0,-k_3,k_2)\label{eq:5.27}
\end{equation}
\begin{equation}
  \vec \xi_2(\vec k)=\frac{1}{wp}(p^2,i\partial_xk_2,i\partial_xk_3).\label{eq:5.28}
\end{equation}


\section{\label{sec:force}Casimir force for plain geometry}


Let us assume for a moment that the dielectric $A$ covers the half-plain 
$-\infty<x<0$ and dielectric $B$ the region $a<x<R$. Then the force 
per unit square on the surface of $B$ is equal to the difference of 
the stress tensors $T_{11}(a-)-T_{11}(R+)$. The term
$T_{11}(R+)$ vanishes, but $T_{11}(a-)$ contains contributions from
the reflection at $x=R$. These contributions vanish in the limit
$R\to\infty$. Then only the Green's function for arguments between
the two dielectrics contributes, which we have calculated in the last
section. This argument is necessary because otherwise the stress
tensor at infinity gives rise to an infinite ``volume force''.\\
The stress tensor for 1-direction is given by
\begin{eqnarray}
  T_{11}(\vec x) &=& \frac{1}{2}\bigl<E_2(\vec x)^2+E_3(\vec x)^2-E_1(\vec x)^2
  \nonumber\\
  & & {}+B_2(\vec x)^2+B_3(\vec x)^2-B_1(\vec x)^2\bigr>'.\label{eq:6.1}
\end{eqnarray}
To find the Green's functions for the
magnetic field, we use the field equations. From
\begin{equation}
  \vec E=-\partial_t\vec A\quad{\rm and}\quad \vec B=\vec\nabla\wedge\vec A\label{eq:6.4}
\end{equation}
we get for the Fourier transforms
\begin{equation}
  \vec B(k)=\frac{1}{\omega}\vec k\wedge\vec E(k).\label{eq:6.5}
\end{equation}
By (\ref{eq:5.1}) (\ref{eq:5.2}) we have
\begin{equation}
  \frac{1}{\omega}\vec k\wedge\vec\xi_1(\vec k)
  =\xi_2(\vec k)\quad{\rm and}\quad 
  \frac{1}{\omega}\vec k\wedge\vec\xi_2(\vec k)
  =-\xi_1(\vec k),\label{eq:6.6}
\end{equation}
using $k^2=\omega^2$.
Now we are ready to calculate the Casimir force. 
We obtain with (\ref{eq:5.26})
\begin{equation}
  \label{eq:6.8}
  T_{11}(\vec x)
  =\frac{1}{2\beta}\sum_me^{i\omega_m\tau}\frac{1}{(2\pi)^2}
  \int\frac{d^2p}{w^2}\Bigl(w^2+\partial_x\partial_{x'}-p^2\Bigr)
  \Bigl(f_1(w;x,x')+f_2(w;x,x')\Bigr)\,\Big|_{x=x'},
\end{equation}
where the $f_j$ are defined in (\ref{eq:5.26}). Only the terms depending on the
difference $(x-x')$ in $f_j$ contribute because they are proportional to
\begin{equation}
  w^2+s_0^2-p^2=2s_0^2.
\end{equation}
This leads to the following final result
\begin{equation}
  T_{11}=\frac{1}{\beta}\sum_{\omega_m}\frac{1}{(2\pi)^2}\int d^2p\,
  is_0\left(\frac{r_1^2e^{2is_0a}}{1-r_1^2e^{2is_0a}}
      +\frac{r_2^2e^{2is_0a}}{1-r_2^2e^{2is_0a}}\right),\label{eq:6.9}
\end{equation}
where $r_1$ and $r_2$ are the reflection factors for the two modes given by
(\ref{eq:5.10.1}) and (\ref{eq:5.10.2}).
The result (\ref{eq:6.9}) is the Lifshitz formula 
\cite{abrikosov} \cite{lifshits}. 


\section{\label{sec:spherical}Spherical geometry}


As mentioned in Sect.3  for compact dielectrics the energy $E_D$ is always
divergent, since it ignores the self-energy contribution $E_G$ in equation
(\ref{eq:3.21}). Nevertheless the above method to calculate the photon
Green's function still works for spherical geometries as shall be shown in this
section. As a consequence, it is not hard to compute the Casimir force on
the spherical shell and rederive the result of ref. [10].
 
The region occupied by the dielectric is now the sphere
\begin{equation}
  K=\{\vec x\in\bbbr^3\>\vert\>\vec x^2\le R^2\}.\label{eq:7.1}
\end{equation}
The free Green's function (\ref{eq:3.11}) (\ref{eq:3.12}) is then written in spherical
coordinates $(r,\vartheta,\varphi)$ using the expansion
\begin{equation}
  e^{i\vec k\cdot\vec x}=4\pi\sum_{lm}i^lj_l(kr)Y^m_l(\frac{\vec x}{r})
  Y^m_l(\frac{\vec k}{ k})^*,\label{eq:7.2}
\end{equation}
where $j_l$ are spherical Bessel functions and $Y^m_l$ the spherical
harmonics. For $\vec x\ne \vec x\,'$ we then have
\begin{equation}
  D^0_{jj'}(w,\vec x,\vec x\,')
  = \frac{2}{\pi}\sum_{\lambda=1,2}
      \xi_{\lambda j}(\vec x)\xi_{\lambda j'}(\vec x\,')
      \int\limits_0^\infty dk\,\frac{k^4}{k^2-w^2}
  \sum_{lm}j_l(kr)j_l(kr')Y^m_l(\vartheta,\varphi)Y^m_l(\vartheta',\varphi')^*\,.
  \label{eq:7.3}
\end{equation}
The $k$-integral can be extended to $(-\infty,+\infty)$ and then
computed using the theorem of residues, taking $w=i\omega_n$ (\ref{eq:3.10})
into account. The result is
\begin{equation}
  \frac{2}{\pi}\int\limits_0^\infty\cdots
  =iw^3\bigl[\theta(r-r')
   h^{(1)}_l(wr)j_l(wr')
   +\theta(r'-r)j_l(wr)h_l^{(1)}(wr')\bigr],\label{eq:7.4}
\end{equation}
where $h_l^{(1)}$ are the Bessel functions of the third kind.
The two transversal polarization vectors perpendicular to $\vec k\sim
\vec p=-i\partial/\partial\vec x$ are constructed as follows: First we choose
\begin{equation}
  \vec\xi_1(\vec x)=\frac{1}{\sqrt{l(l+1)}}\vec x\wedge\vec p=
  \frac{1}{\sqrt{l(l+1)}}\vec L.\label{eq:7.5}
\end{equation}
This is the angular momentum operator which gives the so-called electric
TE mode; it is normalized according to
\begin{equation}
  \vec\xi_1^2=\frac{1}{ l(l+1)}\vec L^2.\label{eq:7.6}
\end{equation}
The second polarization vector is then
\begin{equation}
  \vec\xi_2(\vec x)=\frac{1}{ w\sqrt{l(l+1)}}\vec p\wedge\vec L.\label{eq:7.7}
\end{equation}
This is the TM mode which is also normalized because $|\vec k|=w$.
Since the metric tensor in spherical coordinates is non-trivial, it is
convenient to distinguish between upper and lower tensor components.
Let $Q=l(l+1)$.
In components we have
\begin{eqnarray}
  \label{eq:7.8}
  \xi_1^i(r,\phi,\theta)
   &=&\frac{1}{\sqrt{Q}}\,\frac{1}{r\sin\theta}\,
      \Big(\,0\,,\,-\partial_\theta\,,\,\partial_\phi\,\Big)\nonumber\\
  \xi_{1\,i}(r,\phi,\theta)
  &=&\frac{1}{\sqrt{Q}}\,r\,
     \Big(\,0\,,\,-\sin\theta\,\partial_\theta\,,\,\frac{1}{\sin\theta}\,\partial_\phi\,\Big)\\
  \label{eq:7.9}
  \xi_2^i(r,\phi,\theta)
  &=&\frac{1}{k\sqrt{Q}}\,
     \Big(\,\frac{Q}{r}\,,\,\frac{\partial_rr}{r^2\sin^2\theta}\,\partial_\phi\,,\,
          \frac{\partial_rr}{r^2}\,\partial_\theta\,\Big)\nonumber\\
  \xi_{2\,i}(r,\phi,\theta)
  &=&\frac{1}{k\sqrt{Q}}\,
     \Big(\,\frac{Q}{r}\,,\,\partial_rr\,\partial_\phi\,,\,\partial_rr\,\partial_\theta\,\Big)
\end{eqnarray}
We have orthogonality in the sense \cite{milton}
\begin{equation}
  \int\Bigl(\xi_1^i(\vec x)f(r)Y^m_l(\frac{\vec x}{ r})\Bigl)^*
  \xi_{1\,i}(\vec x)g(r)Y^{m'}_{l'}(\frac{\vec x}{r})\,d^3x
  = \delta_{ll'}\delta_{mm'}\int\limits_0^\infty dr\,r^2f^*(r)g(r)
\end{equation}
\begin{equation}
  \int\Bigl(\xi_2^i(\vec x)f(r)Y^m_l(\frac{\vec x}{r})\Bigl)^*
  \xi_{2\,i}(\vec x)g(r)Y^{m'}_{l'}(\frac{\vec x}{r})\,d^3x 
  =\delta_{ll'}\delta_{mm'}\frac{1}{ w^2}\!\!\int\limits_0^\infty\!\! dr\, 
   \Bigl[\frac{d}{dr}(rf^*)\frac{d}{dr}(rg)+l(l+1)f^*g\Bigr],\label{eq:7.10}
\end{equation}
where we sum over the lower and upper indices $i$. The mixed expressions are
$0$. 
This has the important consequence that different $l$'s and
polarizations don't mix. Therefore, only the radial integrations are
non-trivial.
The free Green's function (\ref{eq:7.3}) is now equal to
\begin{equation}
  D_j^{0\,j'}(w,\vec x,\vec x\,')=\sum_{\lambda=1,2}\xi_{\lambda j}
  (\vec x)\xi_\lambda^{j'}(\vec x\,')
  \sum_{lm}Y^m_l(\vartheta,\varphi)Y^m_l(\vartheta',\varphi')^*
  D^0(r,r'),\label{eq:7.11}
\end{equation}
where
\begin{equation}
  D^0(r,r')
  =iw^3\Bigl[\theta(r-r')h^{(1)}_l(wr)j_l(wr')
   +\theta(r'-r)h^{(1)}_l(wr')j_l(wr)\Bigl],\label{eq:7.12}
\end{equation}
and for the infinite dielectric we have
\begin{equation}
D^1(w,\vec x,\vec x\,')=\frac{1}{\eps}D^0(\sqrt{\eps}w,\vec x,\vec x\,').
\label{eq:7.13}
\end{equation}
The essential point in the solution of the problem is the calculation of
(\ref{eq:4.16}). For the TE mode and $r,r'<R$ we get
\begin{eqnarray}
  \lefteqn{\theta_KD^1\theta_{\bar K}D^0\theta_K\big|_l(r,r')
    = \int\limits_R^\infty
      d\ro\,\ro^2D^1(r,\ro)D^0(\ro,r')}\nonumber\\
  &&= -\langle h^{(1)}_l,h^{(1)}_l\rangle_1 
      \sqrt{\eps}w^6j_l(\sqrt{\eps} wr)j_l(wr')\,,\label{eq:7.14}
\end{eqnarray}
where we have introduced the non-symmetric bilinear form of Bessel functions:
\begin{eqnarray}
  \langle f_l,g_l\rangle_1
   &=&\int\limits_0^R d\ro\,\ro^2f_l(w\ro)
      g_l(\sqrt{\eps}w\ro) \nonumber\\
   &=&\frac{R}{\alpha w^2} 
      \Bigl[wf'_l(wR)g_l(\sqrt{\eps}wR)-\sqrt{\eps}wf_l(wR)g'_l(\sqrt{\eps}wR)
      \Bigl].
  \label{eq:7.15}
\end{eqnarray}
Then (\ref{eq:4.16}) operates multiplicatively, indeed,
\begin{equation}
  \bigl(\theta_KD^1\theta_{\bar K}D^0\bigr)\theta_KD^1\theta_{\bar K}\big|_l=\gamma_{1l}
  \theta_KD^1\theta_{\bar K}\big|_l,\label{eq:7.16}
\end{equation}
where
\begin{equation}
  \gamma_{1l}=\sqrt{\eps}w^6\langle h_l^{(1)},h_l^{(1)}\rangle_1
  \langle j_l,j_l\rangle_1.\label{eq:7.17}
\end{equation}
Since
\begin{equation}
  \bigl(\theta_KD^1\theta_{\bar K}
  -\alpha\theta_KD^1\theta_{\bar K}D^0\theta_{\bar K}\bigr)\big|_l
  = -\alpha\sqrt{\eps}w^6\langle j_l,h^{(1)}_l\rangle_1j_l(\sqrt{\eps}wr)
  h_l^{(1)}(wr')\,,\label{eq:7.18}
\end{equation}
(see also (\ref{eq:a10}) in the appendix) we obtain for (\ref{eq:4.17})
\begin{eqnarray}
  \theta_KD_j^{\,\,j'}\theta_{\bar K}\big|_{\lambda=1}
  &=&-\frac{1}{ 1+\alpha^2\gamma_1}
     \alpha\sqrt{\eps}w^6\langle j_l,h_l^{(1)}\rangle_1
  \nonumber\\
  & &{}\times \xi_{1\, j}(\vec x)\xi_1^{j'}(\vec x\,')
     Y^m_l(\vartheta,\varphi)Y^m_l(\vartheta',\varphi')
     j_l(\sqrt{\eps}wr) h_l^{(1)}(wr')
  \label{eq:7.19}
\end{eqnarray}
for $\lambda=1$ and similarly for $\lambda=2$
\begin{eqnarray}
  \theta_KD_j^{\,\, j'}\theta_{\bar K}\big|_{\lambda=2}
  &=&-\frac{1}{1+\alpha^2\gamma_2} \alpha\frac{w^2}{\sqrt{\eps}}
     \langle j_l,h_l^{(1)}\rangle_2
  \nonumber\\
  & &{}\times \xi_{2\, j}(\vec x)\xi_2^{j'}(\vec x\,')
     Y^m_l(\vartheta,\varphi)Y^m_l(\vartheta',\varphi')
     j_l(\sqrt{\eps}wr) h_l^{(1)}(wr')\,,
  \label{eq:7.20}
\end{eqnarray}
with
\begin{equation}
  \label{eq:7.21}
  \gamma_2=\frac{w^2}{\sqrt{\eps}} 
    \langle j_l,j_l\rangle_2\langle h^{(1)}_l,h^{(1)}_l\rangle_2,
\end{equation}
where we have introduced a second bilinear form for Bessel functions
\begin{eqnarray}
  \langle f_l,g_l\rangle_2
  &=&\int\limits_0^R dr\,
     \Bigl[\frac{d}{ dr}\Bigl(rf_l(wr)\Bigl)\frac{d}{ dr}
           \Bigl(rg_l(\sqrt{\eps}wr)\Bigl)
           +l(l+1)f_l(wr)g_l(\sqrt{\eps}wr)
     \Bigl]\nonumber\\
  &=&\Bigl[\frac{d}{dR}\Bigl(Rf_l(wR)\Bigl)\Bigr]Rg_l(\sqrt{\eps}wR)
      +w^2\langle f_l,g_l\rangle_1.\label{eq:7.22}
\end{eqnarray}
For (\ref{eq:4.18}) we obtain:
\begin{equation}
  \theta_{\bar K}D_j^{\,\,j'}\theta_{\bar K}\big|_\lambda
  = D_j^{0\,j'}\big|_\lambda+\sum_{lm}\mu_{\lambda l} iw^3
    \xi_{\lambda j}(\vec x)\xi_\lambda^{j'}(\vec x\,')
    Y^m_l(\vartheta,\varphi)Y^m_l(\vartheta',\varphi')h_l^{(1)}(wr)h_l^{(1)}(wr')\,,
  \label{eq:7.23}
\end{equation}
where for $\lambda=1$
\begin{equation}
  \mu_{1l}=-\frac{\alpha^2\sqrt{\eps}w^6\langle j_l,j_l\rangle_1
    \langle j_l,h^{(1)}_l\rangle_1}{1+\alpha^2\sqrt{\eps}w^6 
    \langle j_l,j_l\rangle_1\langle h^{(1)}_l,h^{(1)}_l\rangle_1},\label{eq:7.24}
\end{equation}
and similarly for $\lambda=2$:
\begin{equation}
  \mu_{2l}=-\frac{\alpha^2\frac{w^2}{\sqrt{\eps}}\langle j_l,j_l\rangle_2
    \langle j_l,h^{(1)}_l\rangle_2}{1+\alpha^2\frac{w^2}{\sqrt{\eps}} 
    \langle j_l,j_l\rangle_2\langle h^{(1)}_l,h^{(1)}_l\rangle_2}.\label{eq:7.25}
\end{equation}


\section{\label{sec:metallic}Comments on the metallic limit}


In this section the metallic limit $\eps\to0$ is considered. By the
Lorentz-Lorenz relation (\ref{eq:3.24}) this means to put the atomic
polarization $\alpha_0=3$. It is not clear whether in our microscopic 
theory the
perturbation expansion for the Green's function converges for $\alpha_0=3$.
Our microscopic theory does not give an adequate description for
metals. In fact we'll show below that in infinite homogeneous dielectrics the
series only converges for $\alpha_0<\frac{3}{2}$. Nevertheless for the Casimir
effect only Green's function that crosses the boundaries contributes to the
force. 
Indeed the series then converges in the TE mode, at least, for small frequencies $p^2>2u^2$, where
$u=w_n=-iw$. We'll show this only for the special case of flat geometries. 

The Green's function of the radiation field is given by
\begin{equation}
  \label{eq:d1}
  D=D^0+\alpha_0D^0\theta_AD^0+\alpha_0^2D^0\theta_AD'\theta_AD^0\,,
\end{equation}
where 
\begin{equation}
  \label{eq:d1.1}
  D'=D_0'+\alpha_0D_0'\theta_AD_0'+\alpha_0^2\left(D_0'\theta_A\right)^2D_0'+\dots
\end{equation}
with $D_0'=D^0+\frac{1}{3}$. In momentum-space $D_0'$ is given by
\begin{eqnarray}
  \label{eq:d2}
  D_0'(iu,\vec k)&=&\frac{-u^2\delta_{ij}-k_ik_j}{k^2+u^2}
        +\frac{1}{ 3}\delta_{ij}\nonumber\\
      &=&\frac{1}{3}\,\frac{-2u^2+k^2}{k^2+u^2}
         \,\left(\delta_{ij}-\frac{k_ik_j }{ k^2}\right)
         - \frac{2}{3}\frac{k_ik_j}{ k^2}
\end{eqnarray}
It is clear that in infinite homogeneous dielectrics the perturbation
expansion only converges for $\alpha_0<\frac{3}{2}$ because of the
dipole-dipole interaction. 
 
For the Green's function that crosses the boundaries, let us first consider
the case $u=0$. Then only the dipole-dipole interaction contributes. We have
\begin{equation}
  \label{eq:d3}
  D_0'(0,\vec k)=-\frac{k_ik_j}{k^2}+\frac{1}{3}\delta_{ij}
  =D_0^l(\vec k)+\frac{1}{3}\delta_{ij} 
\end{equation}
Let $A$ be the region of the right half-plain with $0<x<+\infty$. The
longitudinal Green's function is given by 
\begin{equation}
  \label{eq:d4}
  D_0^l(x-y)=-(-i\partial_x,p_2,p_3)^T(i\partial_y,p_2,p_3)d^0(x-y)\,,
\end{equation}
where
\begin{equation}
  \label{eq:d5}
  d^0(x)=\frac{1}{2\pi}\int dk_1\frac{1}{k_1^2+p^2}e^{ik_1x}
        =\frac{1}{2p}\left(\theta(x)e^{-px}+\theta(-x)e^{px}\right)\,,
\end{equation}
with $p=\sqrt{p_2^2+p_3^2}$.
Then a straight-forward
calculation gives
\begin{equation}
  \label{eq:d6}
  \theta_AD_0^l\theta_AD_0^l\theta_{\bar A}
  =-\frac{1}{2}\theta_AD_0^l\theta_{\bar A}
\end{equation}
For the Green's function $\theta_AD\theta_{\bar A}$ we get
\begin{equation}
  \label{eq:d7}
  \theta_AD\theta_{\bar A}
  =\bigl(1-\frac{\alpha_0}{2}(1-\frac{\alpha_0}{6}+(\frac{\alpha_0}{6})^2+\dots)
   \bigr)\theta_AD_0^l\theta_{\bar A}
\end{equation}
This is convergent for $\alpha_0<6$. Hence we get
\begin{equation}
  \label{eq:d8}
   \theta_AD\theta_{\bar A}
   =\bigl(1-\frac{\alpha_0}{2}\frac{1}{1+\frac{\alpha_0}{6}}\bigr)
    \theta_AD_0^l\theta_{\bar A}
\end{equation}
This is $0$ for $\alpha_0=3$ as it should for metals. For the reflection we
get with 
$\bigl(\theta_{\bar A}D_0^l\theta_AD_0^l\theta_{\bar A}\bigr)(x,x')
 =-\frac{1}{2}D_0^l(x+x')$ for $x,x'>0$
\begin{equation}
  \label{eq:d9}
  D(x,x')
  =D_0^l(x-x')-\frac{\alpha_0}{2}\frac{1}{1+\frac{\alpha_0}{6}}D_0^l(x+x')
  =D_0^l(x-x')-\frac{\eps-1}{\eps+1}D_0^l(x+x')
\end{equation}
where
\begin{equation}
  \label{eq:d10}
  \eps=1+\frac{\alpha_0}{1-\frac{\alpha_0}{3}}\,.
\end{equation}

For $u>0$ it is much harder to show that series converges in the metallic
limit. We will show this only for the special case where $p^2>2u^2>0$ and only
for TE mode. Let the atomic polarization be any function with
\begin{equation}
  \label{eq:d11}
  \alpha_0(iu)<3 \quad\text{for}\quad u>0  \quad,\quad\alpha_0(0)=3\,.
\end{equation}
We express the expansion for the Green's function $D'$ (\ref{eq:d1.1})
formally by the integral equation
\begin{equation}
  \label{eq:d12}
  D'=D_0'+\alpha_0D_0'\theta_AD'\,,
\end{equation}
where the TE mode of the free Green's function $D_0'$ is given by
\begin{equation}
  \label{eq:d13}
  D_0'(iu,\vec k)=D^0(u,\vec k)+\frac{1}{3}
  =\frac{-u^2}{k^2+u^2}+\frac{1}{3}\,.
\end{equation}
For $p^2=k_2^2+k_3^2>2u^2>0$ the expansion for Green's function $D_1'$ in a
infinite homogeneous dielectric converges and is given by
\begin{equation}
  \label{eq:d14}
  D_1'(iu,\vec k)=\frac{\alpha}{\alpha_0}+\frac{\alpha^2}{\alpha_0^2}D^1(u,\vec k)\,,
\end{equation}
with 
\begin{equation}
  \label{eq:d15}
  D^1(iu,\vec k)=\frac{-u^2}{k^2+\eps u^2}\,.
\end{equation}
This leads to the complementary integral equation
\begin{equation}
  \label{eq:d16}
  D'=D_1'-\alpha_0D_1'\theta_{\bar A} D'\,.
\end{equation}
We then get
\begin{eqnarray}
  \label{eq:d17}
  \theta_AD'\theta_{\bar A}&=&\theta_AD^0\theta_{\bar A} 
  + \alpha_0\theta_AD^0\theta_AD'\theta_{\bar A}
    +\frac{\alpha_0}{3}\theta_AD'\theta_{\bar A}\nonumber\\
  &=&\theta_AD^0\theta_{\bar A} + \alpha_0\theta_AD^0D'\theta_{\bar A}
     -\alpha_0\theta_AD^0\theta_{\bar A}D'\theta_{\bar A}
     +\frac{\alpha_0}{3}\theta_AD'\theta_{\bar A}\,.
\end{eqnarray}
Substitution of $D'$ (149) into the second term of the second line 
and use of 
$D^0+\alpha_0D^0D_1'=\frac{\alpha}{\alpha_0}D^1$ gives
\begin{eqnarray}
  \label{eq:d18}
  \theta_AD'\theta_{\bar A}&=&\frac{\alpha}{\alpha_0}\theta_AD^1\theta_{\bar A} 
    - \alpha\theta_AD^1\theta_{\bar A}D'\theta_{\bar A}
    +\frac{\alpha_0}{3}\theta_AD'\theta_{\bar A}\nonumber\\
  &=&\frac{\alpha}{\alpha_0}\theta_AD^1\theta_{\bar A} 
    - \alpha\theta_AD^1\theta_{\bar A}D^0\theta_{\bar A}
    -\alpha_0\alpha\theta_AD^1\theta_{\bar A}D^0\theta_AD'\theta_{\bar A}
    +\frac{\alpha_0}{3}\theta_AD'\theta_{\bar A} 
\end{eqnarray}
Now the operator 
\begin{equation}
  \label{eq:d19}
  \bigl(-\alpha_0\alpha\theta_AD^1\theta_{\bar A}D^0\theta_A
  +\frac{\alpha_0}{3}\bigr)\theta_AD^1\theta_{\bar A}
  =\gamma\theta_AD^1\theta_{\bar A}
\end{equation}
operates by a simple multiplication. From (\ref{eq:a8}) in the appendix we get
\begin{equation}
  \label{eq:d20}
  \gamma=\frac{\alpha_0}{3}-\frac{\alpha_0(s_1-s_0)^2}{\alpha 4s_1s_0}<1
\end{equation}
for $u>0$, hence the perturbation expansion converges. The reflection factor for the TE
mode is then given by (\ref{eq:5.10.1})
\begin{equation}
  \label{eq:d21}
  r_1(u,p)=-\frac{s_1-s_0}{s_1+s_0}
  =-\frac{\sqrt{\eps u^2+p^2}-\sqrt{u^2+p^2}}{\sqrt{\eps u^2+p^2}+\sqrt{u^2+p^2}}\,.
\end{equation}

This formula for the reflection for the TE mode only holds for
$p^2>2u^2>0$. For $u=0$ the TE mode does not contribute to the reflection
and to the free energy, independent of the model adopted for the
dielectric function. 

This cannot be deduced directly from the Lifshitz
formula, where the TE mode does not contribute only if $\lim_{u\to
  0}\eps(iu)u^2=0$, which does not hold in the plasma model for metals where
\begin{equation}
  \label{eq:d22}
  \alpha_0(iu)=\frac{u_p^2}{u^2+\frac{u_p^2}{3}}
  \longrightarrow \eps(iu)=1+\frac{u_p^2}{u^2}\,.
\end{equation}
Although $\lim_{u\to0}r_1(u,p)\neq0$ for the plasma model, in our microscopic
theory the TE mode does not contribute to the free energy for $u=0$. 

A different result is also obtained for the TM mode for the reflection at a
finite metallic plate in the region $0>x>-R$. It can be seen from formula
(\ref{eq:d22}), that for $u\to 0$ the atomic polarization $\alpha_0(0)=3$ does
not depend from the plasma frequency $u_p$. This has the consequence that as
can be shown, that the reflection at the plate does not depend on the
thickness $R$ of the plate for $u=0$ and we get formula (\ref{eq:d9}) with
$alpha_0=3$. 
Whereas in the macroscopic theory there
is a contribution from the reflection at $x=R$.

The small
frequency behavior is important for the temperature
dependence of the Casimir force, that is widely discussed in recent papers
\cite{temp0,temp1,temp2,temp3,temp4,temp5,temp6}.

%
%

\begin{appendix}

\section{\label{sec:app1}}

In this Appendix we do the main calculations for the plain geometries. 
All the
calculations are simple since it suffices to calculate surface integrals.
First we have to compute the reflection at one plate. We use formula
(\ref{eq:4.15})
\begin{equation}
  \label{eq:a01}
  \theta_KD\theta_{\bar K}=(1+\alpha^2\theta_KD^1\theta_{\bar K}D^0)^{-1}
  \theta_KD^1\theta_{\bar K}(1-\alpha D^0\theta_{\bar K}).  
\end{equation}
To compute the integral 
\begin{equation}
  \label{eq:a1}
  \int_{\bar K}d^3 y\,
  \theta_K(\vec x)D^1(\vec x -\vec y)D^0(\vec y -\vec z)\theta_K(\vec z)
\end{equation}
we use the differential equations 
\begin{eqnarray}
  \label{eq:a2}
  (\lap +\omega^2)D^0_{ij}(\vec y)
  &=&-(\partial_i\partial_j+\omega^2\delta_{ij})\delta(\vec y)\nonumber\\
  (\lap +\eps\omega^2)D^1_{ij}(\vec y)
  &=&-\frac{1}{\eps}(\partial_i\partial_j+\eps\omega^2\delta_{ij})\delta(\vec y).
\end{eqnarray}
On the r.h.s is a function with point support. Since in (\ref{eq:a1}) $\vec y$
and $\vec x$ resp. $\vec z$ have different support $D^0$ and $D^1$ simply
solve the wave equations. By Green's theorem we get a surface integral
\begin{eqnarray}
  \label{eq:a3}  
  &\displaystyle\int_{\bar K}d^3 y\,
    \theta_K(\vec x)D^1(\vec x -\vec y)D^0(\vec y -\vec z)\theta_K(\vec z)
    = {\displaystyle \frac{1}{\alpha\omega^2}} 
      \displaystyle\int_{\partial K}d^2  \vec\sigma(y)\,
   \theta_K(\vec x)&\nonumber\\
  & \times\left((\vec \nabla_{\vec y} D^1(\vec x-\vec y))\,D^0(\vec y-\vec z)
    -D^1(\vec x-\vec y)\vec \nabla_{\vec y} D^0(\vec y -\vec z) \right)\theta_K(\vec z)&
\end{eqnarray}
where the surface vector $\vec\sigma(\vec y)$ shows inside the region
$K$. Because of  the hight symmetry of the plain or the spherical 
geometries no integral at all has to be computed.\\
For plain geometries and $A$ representing the region $x<0$ we get
\begin{equation}
  \label{eq:a4}
  \theta_AD^1_1\theta_{\bar A}D^0_1\theta_A =
  \frac{i\omega^2(s_1-s_0)}{\alpha\,4s_1s_0}
  \xi_{1i}\xi_{1j}^*\theta(-x)e^{-is_1x}\theta(-x')e^{-is_0x'}
\end{equation}
for the electric TE mode and
\begin{equation}
  \label{eq:a5}
  \theta_AD^1_2\theta_{\bar A}D^0_2\theta_A =
  \frac{i\omega^2(s_1-s_0)(p^2-s_1s_0)}{\alpha\,4s_1s_0\,\sqrt{\eps}\omega^2}
  \xi_{2i}\xi_{2j}^*\theta(-x)e^{-is_1x}\theta(-x')e^{-is_0x'}
\end{equation}
for the magnetic TM mode.
Similarly we get 
\begin{equation}
  \label{eq:a6}
  \theta_AD^1_1\theta_{\bar A}D^0_1\theta_AD^1_1\theta_{\bar A}=
  \frac{(s_1-s_0)^2}{\alpha^2\,4s_1s_0}
  \times\frac{i\omega^2}{2s_1}\xi_{1i}\xi_{1j}^*\theta(-x)\theta(x')e^{-is_1(x-x')}
\end{equation}
for the TE mode and
\begin{equation}
  \label{eq:a7}
  \theta_AD^1_2\theta_{\bar A}D^0_2\theta_AD^1_2\theta_{\bar A}=
  \frac{(s_1-s_0)^2(p^2-s_1s_0)^2}{\alpha^2\eps\omega^4\,4s_1s_0}\times
  \frac{i\omega^2}{2s_1}\xi_{2i}\xi_{2j}^*\theta(-x)\theta(x')e^{-is_1(x-x')}
\end{equation}
for the TM mode.
Hence $\theta_AD^1\theta_{\bar A}D^0\theta_A$ operates by a simple
multiplication on $\theta_AD^1\theta_{\bar A}$ with the factors
\begin{equation}
  \label{eq:a8}
  \gamma_1=\frac{(s_1-s_0)^2}{\alpha^2\,4s_1s_0}\quad\text{and}\quad
  \gamma_2=\frac{(s_1-s_0)^2(p^2-s_1s_0)^2}{\alpha^2\eps\omega^4\,4s_1s_0}
\end{equation}
for the two modes. 
Now we compute formula (\ref{eq:4.17})
\begin{equation}
  \label{eq:a9}
  \theta_AD\theta_{\bar A}=\frac{1}{ 1+\alpha^2\gamma}(\theta_AD^1\theta_{\bar A}
   -\alpha \theta_AD^1\theta_{\bar A}D^0\theta_{\bar A}).
\end{equation}
The expression in the bracket can again be computed by the use of the differential
equations (\ref{eq:a2}). The Dirac functions in the differential equations
eliminates the first part. With Green's theorem we obtain
\begin{eqnarray}
  \label{eq:a10}
  \lefteqn{\theta_AD^1\theta_{\bar A}-\alpha
           \theta_AD^1\theta_{\bar A}D^0\theta_{\bar A}}\nonumber\\
  &=&\frac{\alpha}{\alpha\omega^2}\int_{\bar A}d^3y\theta_A(\vec x)
  \left(\lap_yD^1(\vec x-\vec y)\,D^0(\vec y-\vec x\,') - 
         D^1(\vec x-\vec y)\,\lap_yD^0(\vec y-\vec x\,')
   \right)\theta_{\bar A}(\vec x\,')\nonumber\\
  &=&{}-\frac{1}{ \omega^2}\int_{\partial A}d^2\vec\sigma(\vec
           y)\theta_A(\vec x)
   \left( \vec\nabla_y D^1(\vec x-\vec y)D^0(\vec y-\vec x\,')-
          D^1(\vec x-\vec y)\vec\nabla_y D^0(\vec y-\vec x\,')
   \right)\theta_{\bar A}(\vec x\,')
\end{eqnarray}
For plain geometry we obtain
\begin{equation}
  \label{eq:a11}
  \theta_AD^1_1\theta_{\bar A}(1-D^0_1\theta_{\bar A})=
  \frac{i\omega^2(s_1+s_0)}{4s_1s_0}\xi_{1i}\xi_{1j}^*
  \theta(-x)e^{-is_1x}\theta(x')e^{is_0x'}
\end{equation}
and
\begin{equation}
  \label{eq:a12}
  \theta_AD^1_2\theta_{\bar A}(1-D^0_2\theta_{\bar A})=
  \frac{i\omega^2(s_1+s_0)(p^2+s_1s_2)}{4s_1s_0\sqrt{\eps}\omega^2}\xi_{2i}\xi_{2j}^*
  \theta(-x)e^{-is_1x}\theta(x')e^{is_0x'}.
\end{equation}
The factors in (\ref{eq:a9}) simplifies to
\begin{equation}
  \label{eq:a13}
  \frac{1}{ 1+\alpha^2\gamma_1}=\frac{4s_1s_0}{(s_1+s_0)^2} \quad\text{and}\quad 
  \frac{1}{ 1+\alpha^2\gamma_2}=\frac{4s_1s_0\eps\omega^4}{(s_1+s_0)^2(p^2+s_1s_0)^2}\,,
\end{equation}
where in the second factor
$\eps\omega^4=(s_1^2+p^2)(s_0^2+p^2)$ was used in the calculation.
Hence we get for (\ref{eq:a01})
\begin{equation}
\theta(-x)D^A(x,x')\theta(x')
=\Bigl(\frac{iw^2}{s_1+s_0}\xi_{1i}\xi_{1j}^*
    +\frac{i\sqrt{\eps}w^4}{ (s_1+s_0)(p^2+s_1s_0)}\xi_{2i}\xi_{2j}^*
    \Bigl)\theta(-x)e^{-is_1x}\theta(x')e^{is_0x'}\label{eq:a15}
\end{equation}
For the reflection at one plate we have to compute (\ref{eq:4.18}):
\begin{equation}
  \label{eq:a16}
  \theta_{\bar A}D\theta_{\bar A}=\theta_{\bar A}D^0\theta_{\bar A}+\frac{\alpha}{
  1+\alpha^2\gamma}\theta_{\bar A}D^0\theta_AD^1\theta_{\bar A}(1-\alpha D^0
\theta_{\bar A})\,.
\end{equation}
The further integration adds the factors
$-\frac{s_1-s_0}{2s_0}$ and 
$-\frac{(s_1-s_0)(p^2-s_1s_0)}{2s_0\sqrt{\eps}\omega^2}$.
We obtain the result (\ref{eq:5.10}) for the reflection at one plate:
\begin{equation}
  \theta(x)D^A(x,x')\theta(x')=\theta(x)D^0(x,x')\theta(x')
  +\frac{iw^2}{2s_0}\Bigl(r_1(w,p)\xi_{1i}\xi_{1j}^*
        +r_2(w,p)\xi_{2i}\xi_{2j}^*
  \Bigl)\theta(x)e^{is_0x}\theta(x')e^{is_0x'}\label{eq:a17}
\end{equation}
with the two reflection factors
\begin{equation}
  \label{eq:a17.1}
  r_1=-\frac{s_1-s_0}{s_1+s_0}\quad\text{and}\quad 
  r_2=-\frac{(s_1-s_0)(p^2-s_1s_0)}{(s_1+s_0)(p^2+s_1s_0)}\,.
\end{equation}
The reflection factor $r_2$ for the TM mode simplifies with $s_1^2=\eps
w^2-p^2$ and $s_0^2=w^2-p^2$ to
\begin{equation}
  \label{eq:a17.2}
  r_2=-\frac{s_1-\eps s_0}{s_1+\eps s_0}
\end{equation}
For a plate in the region $B$ with $x>a$ we just have to replace in the
formulae $x$ by $a-x$. We will just need the formula for
$\theta_BD^B\theta_{\bar B}$. From (\ref{eq:a15}) we get
\begin{equation}
  \theta(x-a)D^B(x,x')\theta(a-x')
  =\Bigl(\frac{iw^2}{s_1+s_0}\xi_{1i}\xi_{1j}^*
    +\frac{i\sqrt{\eps}w^4}{ (s_1+s_0)(p^2+s_1s_0)}\xi_{2i}\xi_{2j}^*
    \Bigl)\theta(x-a)e^{is_1(x-a)}\theta(a-x')e^{is_0(a-x')}\,.
    \label{eq:a15.1}
\end{equation}
For the Green's function for two parallel plates we use
formula (\ref{eq:5.22.0})
\begin{eqnarray}
  \label{eq:a5.22.0}
  \theta_CD_\lambda\theta_C&=&\theta_CD_\lambda^0\theta_C+
  \mu_\lambda\alpha\theta_CD_\lambda^0\theta_AD_\lambda^A\theta_C+
  \mu_\lambda\alpha\theta_CD_\lambda^0\theta_BD_\lambda^B\theta_C\nonumber\\
  &&{}+\mu_\lambda\alpha^2
    \theta_CD_\lambda^0\theta_AD_\lambda^A\theta_BD_\lambda^B\theta_C+
  \mu_\lambda\alpha^2
  \theta_CD_\lambda^0\theta_BD_\lambda^B\theta_AD_\lambda^A\theta_C 
\end{eqnarray}
where
\begin{equation}
\mu_\lambda=\frac{1}{1-\alpha^2\gamma_\lambda}.\label{eq:a5.22}
\end{equation}
and $\gamma_\lambda$ is given by
\begin{equation}
(\theta_AD^A_\lambda\theta_BD^B_\lambda)\theta_AD^A_\lambda\theta_B=
\gamma_\lambda\theta_AD^A_\lambda\theta_B\label{eq:a5.21}
\end{equation}
First we compute the expression in the quote in (\ref{eq:a5.21}). For $D^A$ and
$D^B$ similar differential equation holds as in (\ref{eq:a2}), so that we
just have to compute a surface integral giving
\begin{equation}
  \label{eq:a23}
  \theta_AD^A_1\theta_BD^B_1\theta_A=
  \frac{-i(s_1-s_0)w^2}{\alpha(s_1+s_0)^2}e^{i2s_0a}\,
  \xi_1\xi_1^* \theta(-x)e^{-is_1x}\theta(-x')e^{-is_0x'}
\end{equation}
for the TE mode and 
\begin{equation}
  \label{eq:a24}
  \theta_AD^A_2\theta_BD^B_2\theta_A=
  \frac{-i\sqrt{\eps}\omega^4(s_1-s_0)(p^2-s_1s_0)}
       {\alpha(s_1+s_0)^2(p^2+s_1s_0)^2} 
  e^{i2s_0a}\,
   \xi_2\xi_2^*\theta(-x)e^{-is_1x}\theta(-x')e^{-is_0x'}
\end{equation}
for the TM mode. A further integration adds the factors
$\frac{i(s_1-s_0)}{\alpha\omega^2}$ and 
$\frac{i(s_1-s_0)(p^2-s_1s_0)}{\alpha\sqrt{\eps}\omega^4}$, so that we get
for $\gamma_\lambda$ in (\ref{eq:a5.21}) 
\begin{equation}
  \label{eq:a25}
  \alpha^2\,\gamma_1=\frac{(s_1-s_0)^2 }{ (s_1+s_0)^2}e^{i2s_0a}
  \quad\text{and}\quad
  \alpha^2\,\gamma_2=\frac{(s_1-s_0)^2(p^2-s_1s_0)^2 }{
                      (s_1+s_0)^2(p^2+s_1s_0)^2}e^{i2s_0a}  
\end{equation}
where the reflection factors from (\ref{eq:a17.1}) appear again. For
$\mu_\lambda$ we get
\begin{equation}
  \label{eq:a25.1}
  \mu_\lambda=\frac{1}{1-r_\lambda^2e^{2is_0a}}
\end{equation}
Finally we have to compute (\ref{eq:a5.22.0}). The second term is just the
reflection at the plate $A$ and is given (\ref{eq:a17}). The third term is the
reflection at the plate $B$, where $x$ has to be substituted by $a-x$. 
For the forth and fifth terms the further integration adds again the factors
$\frac{i(s_1-s_0)}{\alpha\omega^2}$ and 
$\frac{i(s_1-s_0)(p^2-s_1s_0)}{\alpha\sqrt{\eps}\omega^4}$ to (\ref{eq:a23})
and (\ref{eq:a24}).  
This gives for $x$ and $x'$ between the two plates
\begin{eqnarray}
  \label{eq:a26}
  D_\lambda(x,x')&=&D_\lambda^0(x-x')\nonumber\\
  &&{}+\mu_\lambda\, r_\lambda\,
  \frac{i\omega^2}{2s_0}\,\xi_\lambda\xi_\lambda^*
     \left(e^{is_0(x+x')}+e^{is_0(2a-x-x')}\right)\nonumber\\
  &&{}+\mu_\lambda\,r_\lambda^2\,e^{2is_0a}\, 
  \frac{i\omega^2}{2s_0}\xi_\lambda\xi_\lambda^*
     \left(e^{is_0(x-x')}+e^{-is_0(x-x')}\right)
\end{eqnarray}
for $\lambda=1,2$ and with $\mu_\lambda$ given in (\ref{eq:a25.1}) and the reflection
factors $r_\lambda$ given in (\ref{eq:a17.1}). 

\end{appendix}

\end{document}